\documentclass[
 ,draft            
  ]
  {aipproc}

\layoutstyle{8x11single}

\begin{document}

\title{Paradoxical Way for Losers in a Dating Game}


\author{C. M. Arizmendi}{
  address={Departamento de F\'{\i}sica, Facultad de
Ingenier\'{\i}a, Universidad Nacional de Mar del Plata,\\  Av.
J.B. Justo 4302, 7600 Mar del Plata, Argentina\\} }

\begin{abstract}
We study the dating market decision problem in which men and women
repeatedly go out on dates and learn about each other. We consider
a model for the dating market that takes into account progressive
mutual learning. This model consists of a repeated game in which
agents gain an uncertain payoff from being matched with a
particular person on the other side of the market in each time
period. Players have a list of preferred partners on the other
set. The players that reach higher rank levels on the other set
preferences list have also higher probability to be accepted for
dating. A question can be raised, as considered in this study: Can
the less appreciated players do better? Two different kinds of
dating game are combined "à la Parrondo" to foster the less
attractive players. Optimism seems to be highly recommendable,
especially for losers.
\end{abstract}

\maketitle


\section{Introduction}

Matching problems where the elements of two sets have to be
matched by pairs have broad implications in economic and social
contexts. As possible applications one could think of job seekers
and employers, lodgers and landlords, or simply men and women who
want to date. Standard models of matching in economics \cite{Roth
and Sotomayor(1990)} almost always assume that each agent knows
his or her preferences over the individuals on the other side of
the market. This assumption, generally associated with
neo-classical economics, is too restrictive for many markets, and
some interesting work on matching problems with partial
information has recently been published \cite{Zhang(2001), Laureti
and Zhang(2003)}. Specifically,  perfect information supposition
is very far from being a good approximation for the dating market,
in which men and women repeatedly go out on dates and learn about
each other. Recently, a model for the dating market that takes
into account progressive mutual learning was introduced by Das and
Kamenica \cite{Das and Kamenica(2005)}. This model consists of a
repeated game in which agents gain an uncertain payoff from being
matched with a particular person on the other side of the market
in each time period. The problem is related to bandit problems
\cite{Berry and Fristedt(1985)}, in which an agent must choose
which arm of an n-armed bandit to pull in order to maximize
long-term expected reward, taking into account the tradeoff
between exploring, that is learning more about the reward
distribution for each arm, and exploiting, pulling the arm with
the maximum expected reward. However, in Das and Kamenica model
the arms themselves have agency - they can decide whether to be
pulled or not, or whom to be pulled by, and they themselves
receive rewards based on who the puller is. This motivates Das and
Kamenica formulation of the problem as a "two-sided" bandit
problem.\\

The matching problems describe  systems where two sets of persons
have to be matched pairwise. Players have a list of preferred
partners on the other set. The players that reach higher rank
levels on the other set preferences list have also higher
probability to be accepted for dating.
 An interesting question is: Can the less appreciated
players do better? In other words: Can the usual dating game
losers achieve a better performance?  A possible way to accomplish
this goal may be the phenomenon known in the literature as
Parrondo's paradox \cite{Parrondo}, devised in 1996 by the Spanish
physicist Juan M.R. Parrondo, where the alternation of two fair
(or losing) games can result in a winning game. Although initially
introduced as individual games, multiplayer versions of the
paradox, played by a set of N players, have also been studied
\cite{Toral1,Toral2}. In these collective games, a set of $N$
players are arranged in a ring and each round a player is chosen
randomly to play either game $A$ or $B$. The game $A$ is a simple
coin tossing game, where the player wins or loses one unit of
capital with probabilities $p_A$ and $1-p_A$, respectively in
\cite{Toral1}, and a redistribution process where a player is
chosen randomly to give away one coin of his capital to another
player in \cite{Toral2}. Game $A$ is combined with  game $B$, for
which the winning probability depends on the state (winner/loser)
of the nearest neighbors of the selected player. A player is said
to be a \textsl{winner (loser)} if he has \textsl{won (lost)} his
last game. Recently \cite{Amengual2006} a new version of
collective games was introduced, where besides obtaining the
desired result of a winning game out of two fair games, another
feature appears: the value of the $A/B$ mixing probability
$\gamma$ determines whether you end up with a winning or a losing
game $A + B$. In  each round, a player is selected randomly for
playing. Then, with probabilities $\gamma$ and $1 - \gamma$,
respectively, game $A$ or $B$ is played. Game $A$ is the original
coin tossing game. The winning probabilities in game $B$ depend on
the collective state of all players. More precisely, the winning
probability can have three possible values, determined by the
actual number of winners $w$ within the total number of players N,
in the following way
\begin{equation}
p_B=winning\, probability\,  in\,  game
B\:=\:\left\{\begin{array}{ccr}
p^1_B & if & w > [\frac{2N}{3}],\\
p^2_B& if  & [\frac{N}{3}]<w \leq [\frac{2N}{3}],\\p^3_B & if &
w\leq[\frac{N}{3}],\end{array}\right.
\end{equation}
where the brackets $[x]$ denote the nearest integer to the number
x. The set of values ${p^1_B,p^2_B,p^3_B}$ are determined in order
to give a fair game and depend on the total number of players $N$
\cite{Amengual2006}.

 In this paper we consider a repeated
mixing of two different dating games based on both Das and
Kamenica learning dating model and on the last version
\cite{Amengual2006} of collective Parrondo games to analyze the
possibility that the less attractive players do better. Both
dating games are assumed to be fair. For the sake of clarity let
us imagine that the different games that we call $A$ and $B$ are
played in different places with different rules. In game $A$, the
probability that the man proposal be accepted is modelled by coin
tossing, that is $p_A=1/2$. This does not necessarily mean that
every woman flips actually a coin, but it can be thought that all
the variables not considered to construct the preference list,
such as woman's mood, man's way, or the group size dependance of
females selectivity \cite{Fisman2006} are contributing to the
probability of acceptance $p_A$. On the other hand, the
probability of acceptance for game $B$, $p_B$, depends on the
number of previous winners within all players. This collective
influence may seem not so clear at first sight, but it may be
thought that in $p_B$ collective moods contribute, such as if all
woman's friends are dating, that particular woman is better
disposed to accept man's proposal to go out. This is the well
known {\sl herd behavior} that is present in stock market bubbles
as well as in everyday decision making.
\section{The Model}
There are N men and N women, who interact for T time periods. $v^m
_{j}$ is the value of woman $j$ to every man, and $v^w_{i}$ is the
value of man $i$ to every woman. These values are constant through
time. In each period, a man $i$ is chosen randomly from the N
possible men. The expected $i's$ payoff of dating woman $j$:
\begin{equation}
payoff^m_{i,j}[t] = Q^m_{i,j}[t]*p^m_{i,j}[t], \label{payoffm}
\end{equation}
where $Q^m_{i,j}[t]$ is the man $i's$ estimate of the value of
going out with woman $j$ at time $t$ and $ p^m_{i,j}[t]$ is the
man $i's$ estimate at time $t$ of the probability that woman $j$
will go out with him if he asks her out. In this way man's
decision is based on any prior beliefs and the number of rewards
he has received. Both the expected value on a date with that
particular woman and the probability that she will accept his
offer are taken into account by \eqref{payoffm}. The expected
woman $j's$ payoff of dating man $i$ is:
\begin{equation}
payoff^w_{i,j}[t] =  Q^w_{i,j}[t], \label{payoffw}
\end{equation}
where $ Q^w_{i,j}[t]$ is the woman $j's$ estimate of the value of
going out with man $i$ at time $t$. No probability is considered
because  man $i$ considered as a date to be must propose to the
woman $j$. Since the underlying $v^m _{j}$ and $v^w_{i}$ are
constant we define $ Q^m_{i,j}[t]$  as man $i$'s sample mean at
time $t$ of the payoff of going out with woman $j$:
\begin{equation}
Q^m_{i,j}[t]=\sum (v^m _{j} + \epsilon),
\end{equation}
where the sum is made on the effective dates between $i$ and $j$
and $\epsilon$ is noise drawn from a normal distribution.
 In the same way, $ Q^w_{i,j}[t]$ is
woman $j$'s sample mean at time $t$ of the payoff of going out
with man $i$:
\begin{equation}
Q^w_{i,j}[t]=\sum (v^w _{i} + \epsilon).
\end{equation}
  In order to deal with the nonstationarity  of
$p^m_{i,j}[t]$'s, on the other hand, we use a fixed learning rate
for updating the probabilities which allows agents to forget the
past more quickly:
\begin{equation}
p^m_{i,j} [t] = (1 - \eta)p^m_{i,j}[t - 1] + \eta
\:\frac{offers_{i,j} \, accepted[t-1]}{offers_{i,j} \,
made[t-1]},\label{p}
\end{equation}
where $\eta$ is a constant parameter.
\subsection{The Man's Decision Problem}
The top ranked woman from the list of preferred partners of $i$ is
selected to ask out for a date. The
 rank levels of the preference
list are distributed according to the expected $i's$ payoff of
dating woman $j$ \eqref{payoffm}. The man $i$ acts in a greedy way
asking out woman $j$ at the top of his preference list.

\subsection{The Woman's Decision Problem}
The
 rank levels of the women preference
lists are distributed according to the expected woman $j's$ payoff
of dating man $i$ \eqref{payoffw}.
 The woman's
decision problem depends on the game:

\subsubsection{Game $A$}

In both games women have to consider the exploration-exploitation
tradeoff. Exploitation means maximizing expected reward (greedy
choice). Exploration happens when the player selects an action
with lower expected payoff in the present in order to learn and
increase future rewards. One of the simplest techniques used for
bandit problems is the so-called $\epsilon$-greedy algorithm. This
algorithm selects the arm with highest expected value with
probability $1-\epsilon$ and otherwise selects a random arm. We
will use  slightly changed versions of the $\epsilon$-greedy
algorithm in both games. In game $A$ the exploration-exploitation
tradeoff depends on coin tossing, that is the woman accepts the
man's $i$ offer to date with probability $p_A=1/2$ (exploration)
or she acts greedily and goes out with her best $payoff_w$ choice
with probability $1-p_A=1/2$.

\subsubsection{Game $B$}

In game $B$ the choice of exploration or greedy behavior depends
on the collective state of all men players. A man player is said
to be a winner or a loser when he got his date or not,
respectively, in his last game. More precisely, the winning or
exploration probability can have three possible values, determined
by the actual number of winners $w$ within the total number of
players $N$, in the following way

\begin{equation}
p_B=exploration\, probability\,  in\,  game
B\:=\:\left\{\begin{array}{ccr}
p^1_B & if & w > [\frac{2N}{3}],\\
p^2_B& if  & [\frac{N}{3}]<w \leq [\frac{2N}{3}],\\p^3_B & if &
w\leq[\frac{N}{3}],\end{array}\right.\label{pB}
\end{equation}
where the brackets $[x]$ denote the nearest integer to the number
x. The woman accepts the man's $i$ offer to date with probability
$p_B$ (exploration) or she goes out with her best $payoff_w$
choice with probability $1-p_B$. The set of values
${p^1_B,p^2_B,p^3_B}$ are determined in order to give a fair game
and depend on the total number of players $N$ \cite{Amengual2006}.

\begin{figure}[tbp]
\begin{minipage}[c]{.5\textwidth}
\includegraphics[width=\textwidth]{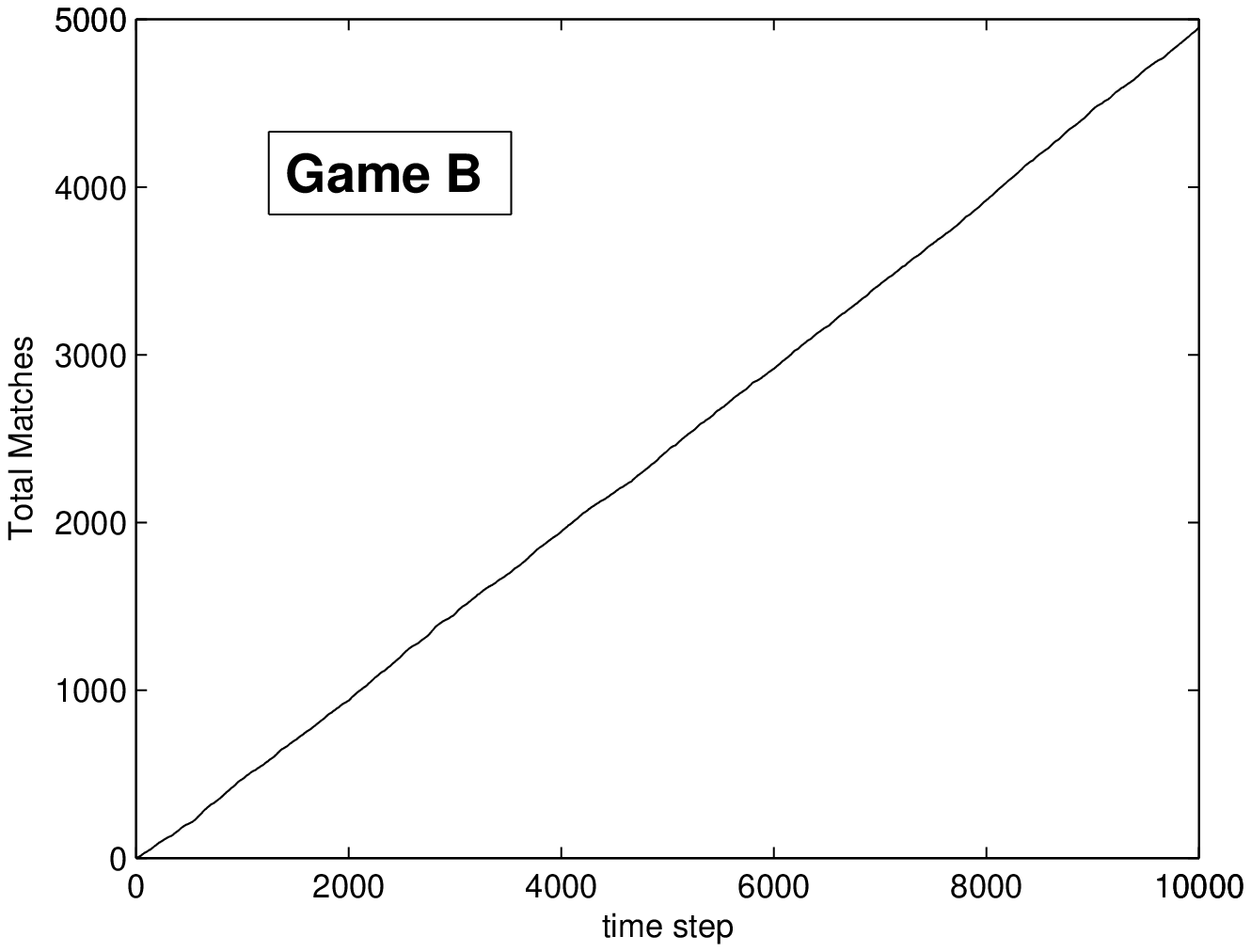}
\end{minipage}%
\begin{minipage}[c]{.5\textwidth}
\includegraphics[width=\textwidth]{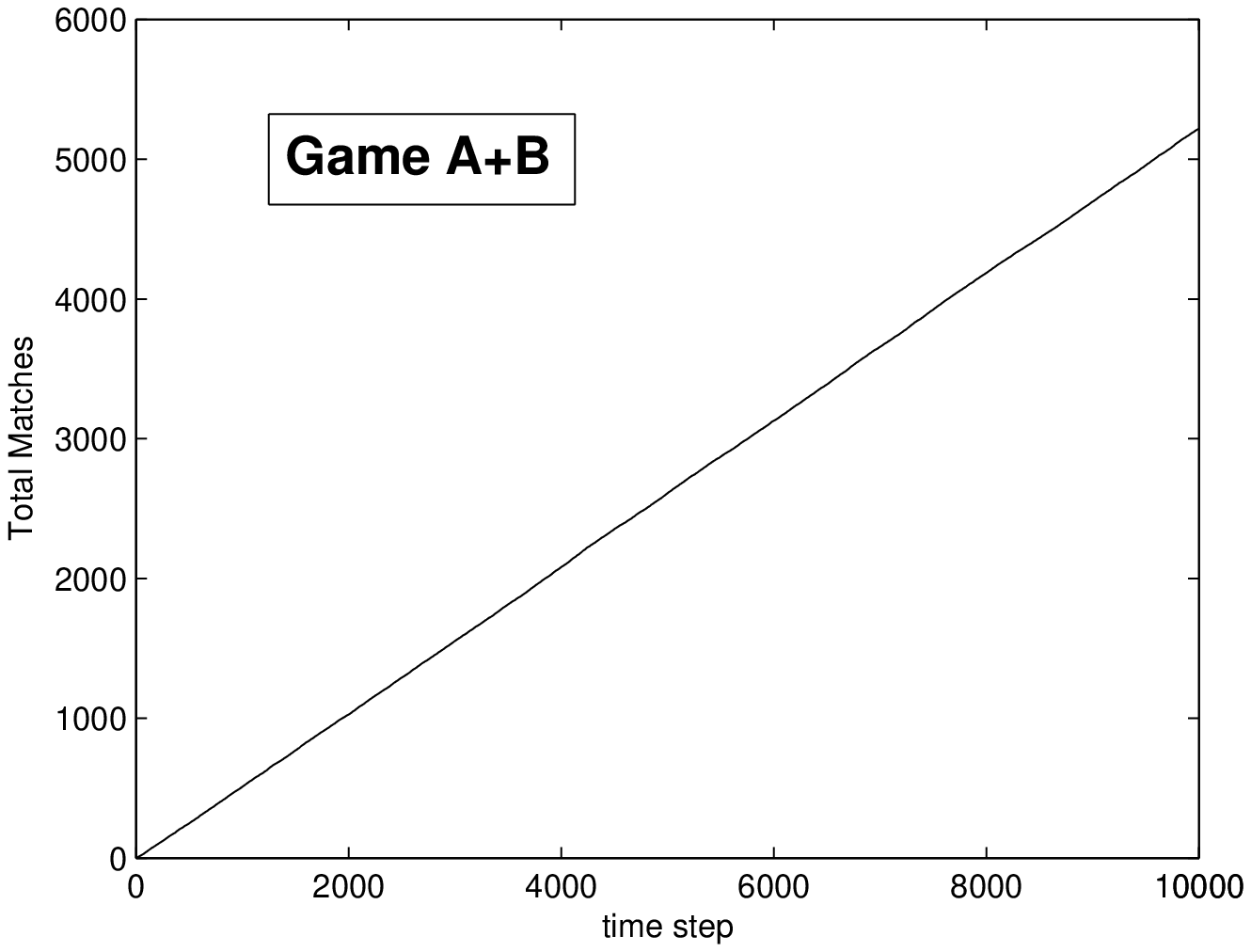}
\end{minipage}
\caption{Time evolution of total number of offers accepted between
dating game $B$ played alone and games $AB$ switched with mixing
probability $\gamma = 1/2$. Total matches vs. time (a) Game $B$;
(b) Games $AB$}
\end{figure}

\section{Results}

Our simulations involve a market of $N=4$ men and $N=4$ women. The
learning rate of probability $p^m_{i,j} [t]$ is $\eta=0.05$. The
set of probabilities for game $B$, ${p^1_B,p^2_B,p^3_B}$
determined to give a fair game \cite{Amengual2006} for $N=4$ are
$p^1_B=0.79$, $p^2_B=0.65$ and $p^3_B=0.15$. The noise signal is
drawn from a normal distribution of standard deviation $0.5$. $v^m
$'s and $v^w$'s are:
\begin{equation}
v^m _{k}=v^w_{k}=N-k+6,
\end{equation}
where $1 \leq k \leq 4$. Reported results are obtained with $1000$
simulations averages on $10^4$ time steps.

\begin{figure}[tbp]
\begin{minipage}[c]{.5\textwidth}
\includegraphics[width=\textwidth]{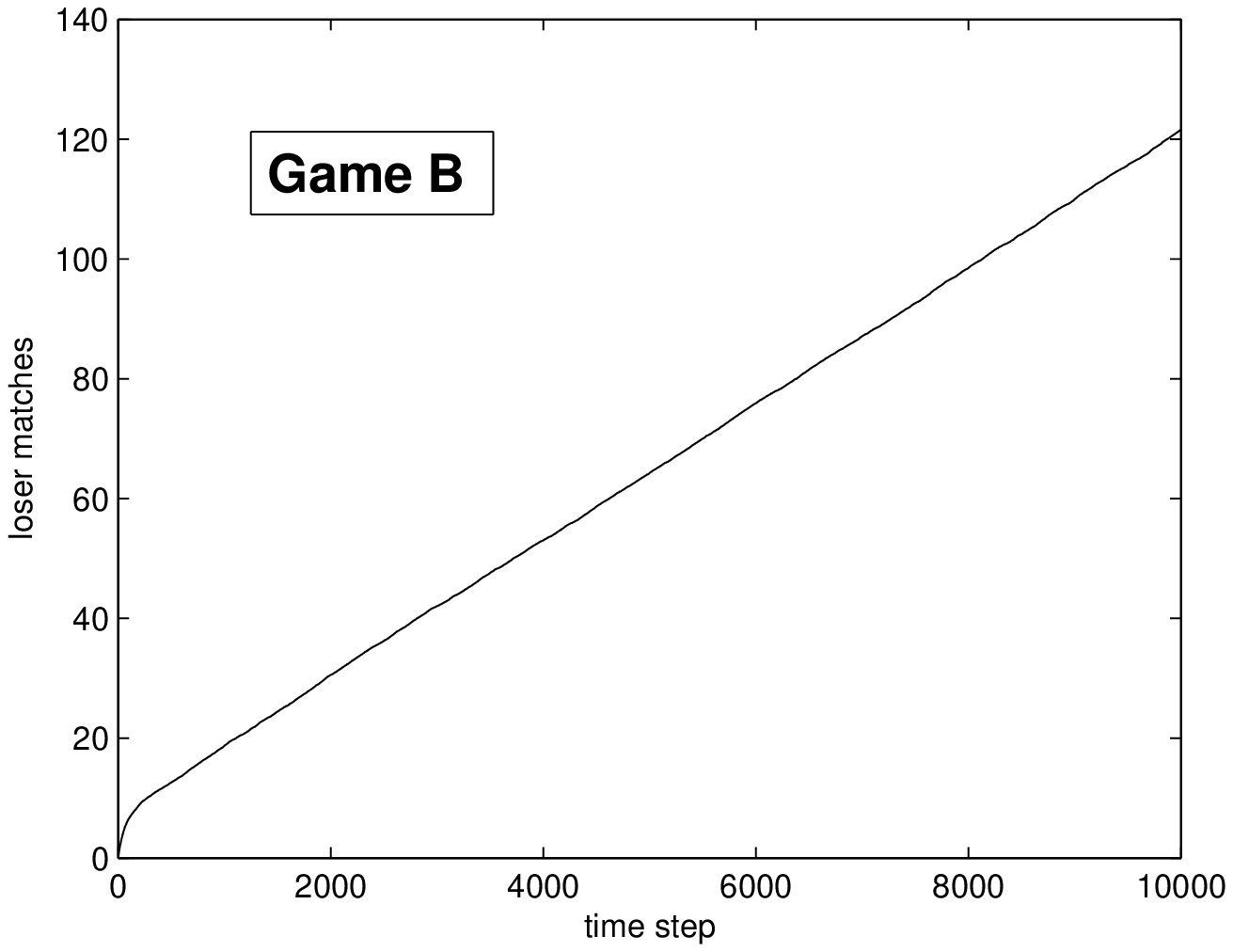}
\end{minipage}%
\begin{minipage}[c]{.5\textwidth}
\includegraphics[width=\textwidth]{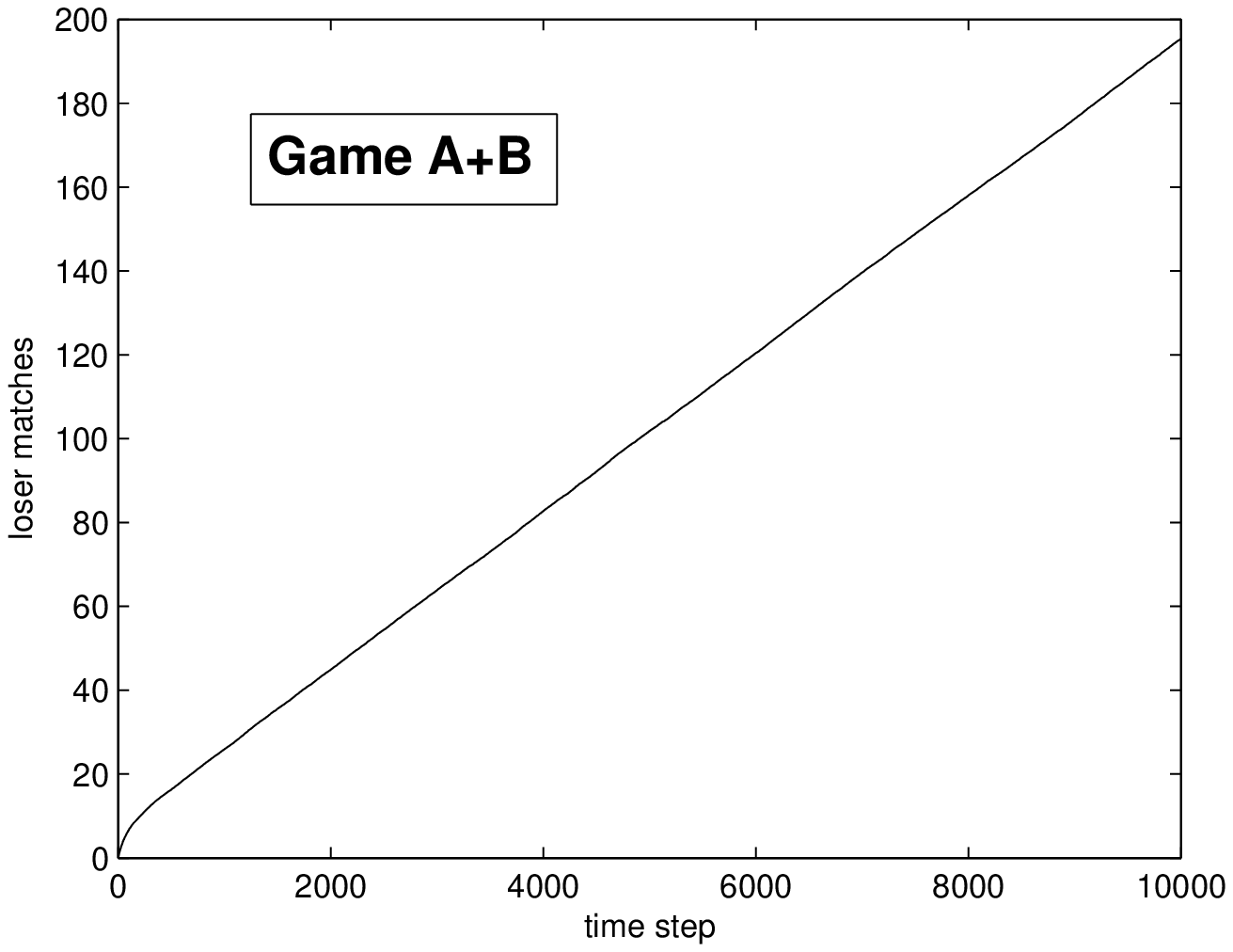}
\end{minipage}
\caption{Time evolution of number of accepted offers from the last
of preferences list (loser) between dating game B played alone and
games AB switched with mixing probability $\gamma = 1/2$. Loser
matches vs. time (a) Game B; (b) Games AB}
\end{figure}

\subsection{Parrondo's Paradox}

 All players benefit
from Parrondo mixing of both games $A$ and $B$ as can be observed
in Fig. 1. In this Figure the total matches, that is the number of
accepted dating offers from all players, is shown as a function of
time for game $B$ played alone and games $A$ and $B$ switched with
mixing probability $\gamma = 1/2$. In the first case, the fairness
of game $B$ is verified when half of all attempts produce a match.
When $A+B$ are played, the paradox produces that more than half of
all attempts are successful, as
can be seen in Fig. 1b. \\
On the other hand, less favored players, i.e. the lowest $v^w$
ones, have an evolution that is shown in Fig. 2, for game $B$
played alone and games $A$ and $B$ switched with mixing
probability $\gamma = 1/2$. In this Figure the advantage of
playing $A+B$ over $B$ for losers can be appreciated. The
comparison with game $A$ is esentially the same for all the
results studied.

\subsection{Optimistic Results}

Das and Kamenica propose as an alternative method to obtain
asymptotic stability in their model for dating couples to suppose
that players are initially optimistic and their level of optimism
declines over time. As they say ``This is another form of patience
- a willingness to wait for the best - and it should lead to more
stable outcomes" \cite{Das and Kamenica(2005)}. A systematic
overestimate of the probability that the dating offer will be
accepted is used to represent
optimism. \\
 Let us  analyze
 optimistic players performance at our dating games model. At time $t$ optimistic players use the
probability estimate:
\begin{equation}
p'^m_{i,j} [t] = \alpha(t) + (1-\alpha(t))p^m_{i,j} [t],
\end{equation}
where $p^m_{i,j} [t]$ is updated as before by \eqref{p} and
$\alpha(t)=(T-t)/T$, with $T$ the total number of time steps in
simulations. Figure 3 shows the evolution of loser accepted offers
 corresponding to optimistic players. The order of loser
 acceptance increase by a factor $10$ comparing optimistic (Fig. 3) and
 non-optimistic (Fig. 2) losers. On the other hand, the advantage of
playing $A+B$ over $B$ for losers is conserved.


\begin{figure}[tbp]
\begin{minipage}[c]{.5\textwidth}
\includegraphics[width=\textwidth]{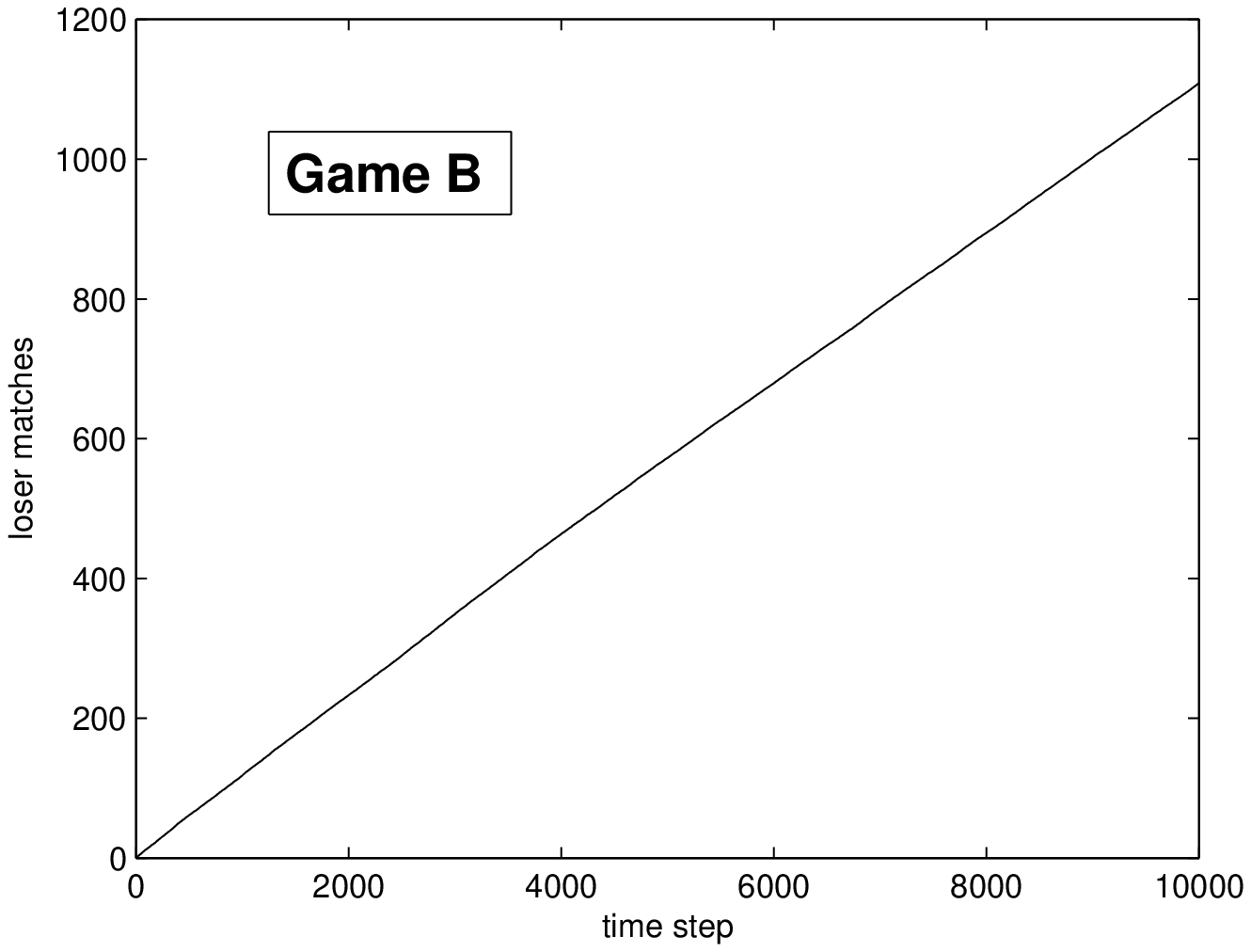}
\end{minipage}%
\begin{minipage}[c]{.5\textwidth}
\includegraphics[width=\textwidth]{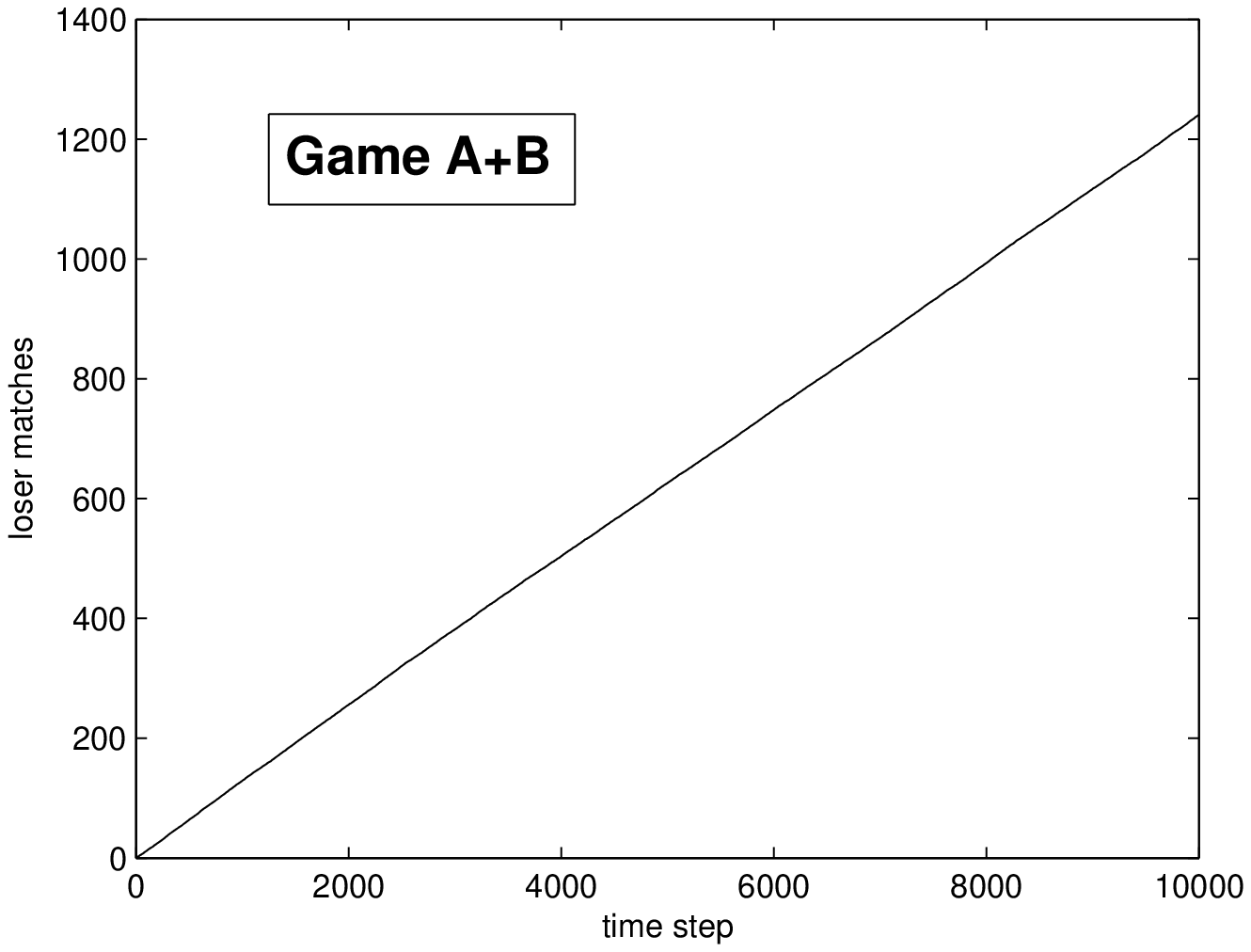}
\end{minipage}
\caption{Optimistic Players: Time evolution of number of accepted
offers from the last of preferences list (loser) between dating
game B played alone and games AB switched with mixing probability
$\gamma = 1/2$. Loser matches vs. time (a) Game B; (b) Games AB}
\end{figure}

\subsection{$N$ and $\gamma$ dependance}
The results are highly dependent on both, the number $N$ of
players and the mixing probability $\gamma$ of the games $A$ and
$B$. As thoroughly explained on \cite{Amengual2006}, the losing or
winning character of the mixed $A+B$ games depends on $N$ and
$\gamma$. We will present in a more extensive way the dependance
on  the number $N$ of players and the mixing probability $\gamma$
of the games $A$ and $B$ elsewhere.

\section{Conclusions}

We  find a way for less qualified dating game players to improve
their performance by means of  a repeated mixing of two different
dating games $A$ and $B$ based on a recent dating market model
\cite{Das and Kamenica(2005)} and on a recent collective game
version of Parrondo's paradox \cite{Amengual2006}.  In game $A$,
the probability associated to exploration-exploitation tradeoff,
that the man proposal be accepted is modelled by coin tossing,
that is $p_A=1/2$.  On the other hand, the probability of
acceptance for game  $B$ depends on the collective state of the
men set obtained through the number of previous winners within all
players. We show that losers benefit from Parrondo mixing of both
games $A$ and $B$. In the optimistic version of our model, when it
is assumed that players are initially optimistic and their level
of optimism declines over time,  loser
 acceptance increase by a factor $10$ and the \textsl {paradoxical} advantage of
playing $A+B$ over $B$ or $A$ for losers is conserved. The results
are highly dependent on both, the number $N$ of players and the
mixing probability $\gamma$ of the games $A$ and $B$.

\begin{theacknowledgments}
  We would like to thank Hugo Lopez Montenegro for many
  interesting and inspiring discussions at the beginning of this
  work. This research was partially supported by Universidad Nacional de
Mar del Plata and ANPCyT (PICTO 11-495).
\end{theacknowledgments}








\end{document}